\documentclass[11pt,3p]{elsarticle}
\makeatletter
\def\ps@pprintTitle{\let\@oddhead\@empty
  \let\@evenhead\@empty
  \def\@oddfoot{\reset@font\hfil\thepage\hfil}
  \let\@evenfoot\@oddfoot
}
\makeatother

\usepackage{amsmath}
\usepackage{amsfonts}
\usepackage{amssymb}
\usepackage{bbm}
\usepackage{amsfonts}
\usepackage{amsthm}

\usepackage{amscd}
\usepackage{mathtools}
\usepackage{mathrsfs}
\usepackage{subcaption}
\usepackage{commath}
\usepackage{lmodern}
\usepackage{color}
\usepackage{a4wide}
\usepackage{bm} 
\usepackage{physics}
\usepackage{enumitem}  
\usepackage{bbm}
\usepackage{booktabs}
\usepackage{hyperref}
\usepackage{float}
\usepackage{graphicx}
\usepackage{svg} 
\hypersetup{ colorlinks=true, urlcolor  = blue, linkcolor = blue,
citecolor = blue1,}
\usepackage{comment}
\usepackage{orcidlink}
\usepackage[english]{babel}
\usepackage[IL2]{fontenc}
\usepackage[utf8]{inputenc}

\usepackage{tabularx}
\usepackage{array}

\usepackage{booktabs}
\usepackage{multirow}
\usepackage{graphicx}

\usepackage[linesnumbered,ruled,vlined]{algorithm2e}
\usepackage[noend]{algpseudocode}

\usepackage{amsmath,amssymb,mathtools,bm}
\usepackage{enumitem}
\usepackage{physics}

\SetKwInput{KwInput}{Input}                
\SetKwInput{KwOutput}{Output}              

\SetCommentSty{mycommfont}

\theoremstyle{definition}

\theoremstyle{remark}
\theoremstyle{remark}

\usepackage[dvipsnames]{xcolor}

\usepackage{ulem}

\begin{document}

\begin{frontmatter}
\title{Interaction order controls stochastic-resonance enhancement and redistribution in delayed multiplex neural networks}
\author[1]{Alina Schlabritz \orcidlink{0009-0008-7678-0089}}
\ead{alina.schlabritz@fau.de}
\author[1,2]{Marius E. Yamakou  \orcidlink{0000-0002-2809-1739}}
\ead{marius.yamakou@fau.de}
\address[1]{Department of Data Science, Friedrich-Alexander-Universit\"at Erlangen-N\"urnberg, N\"urnberger Str. 74, 91052 Erlangen, Germany}
\address[2]{Department of Mathematics, Friedrich-Alexander-Universit\"at Erlangen-N\"urnberg, Cauer Str. 11, 91058 Erlangen, Germany}


\begin{abstract}
We investigate stochastic resonance in a hierarchy of delayed excitable systems, from a single FitzHugh--Nagumo neuron to single-layer and duplex networks with autaptic, pairwise, triadic higher-order, and interlayer interactions.
We show that the order and organization of interactions generate distinct resonance regimes. 
Among the autapse-free single-layer architectures, pairwise and triadic coupling attain comparable maximal responses, but over different ranges of nominal coupling strength: the pairwise enhancement persists across the investigated coupling range, whereas the triadic enhancement is confined to weak coupling. Under the matched participation degrees used here, this difference is largely attributable to the doubled effective coupling generated by the hyperedge construction.
Delayed autaptic feedback primarily shifts the optimal noise level and, within the configurations investigated, yields a genuine enhancement only when pairwise and triadic interactions coexist, revealing a non-additive coupling effect. In duplex networks, interlayer coupling acts predominantly as a response-equalization mechanism: it enhances the weaker layer, generally at the expense of the stronger one, with the redistribution controlled mainly by the uncoupled resonance-capacity gap. Across the sampled duplex configurations, increasing the interlayer delay largely suppresses the response within the investigated delay range. These results identify interaction order, resonance mismatch, and delay as key control parameters for noise-assisted signal processing in multilayer excitable systems.
\end{abstract}

\end{frontmatter}

\section{Introduction}
\label{sec:introduction}

Noise is conventionally regarded as a source of fluctuations that degrades signal detection and transmission. In nonlinear systems, however, stochastic fluctuations can couple constructively to the underlying deterministic dynamics and enhance the response to weak external forcing. A paradigmatic manifestation of this mechanism is stochastic resonance (SR), in which the response to a weak periodic signal becomes maximal at a finite, intermediate noise intensity \cite{gammaitoni-1998}. Originally introduced in connection with noise-driven transitions in models of glacial cycles \cite{sr-iceage}, SR has since been established as a generic nonlinear phenomenon in a broad class of physical and biological systems \cite{wiesenfeld, douglass1993, fauve19835}.

Excitable systems provide a particularly natural setting for SR. In the deterministic excitable regime, a stable resting state coexists with a finite excitation threshold, so that a sufficiently weak periodic perturbation generates only subthreshold oscillations. Stochastic fluctuations can induce threshold crossings, while the periodic forcing biases their occurrence toward preferred phases of the drive. At an intermediate noise intensity, the competition between activation and loss of phase coherence produces an optimal response \cite{good-noise}. This general interplay between excitability, stochastic forcing, and intrinsic timescales also underlies related noise-induced phenomena such as coherence resonance \cite{cr-review, yamakouControlCoherenceResonance2019}, self-induced stochastic resonance \cite{sisr}, inverse stochastic resonance \cite{gutkin-isr}, and diversity-induced resonance and decoherence \cite{diversity-induced-resonance,
diversity-induced-decoherence}. In neuronal systems, these mechanisms influence spike regularity, weak-signal detection, information transmission, and collective synchronization. The FitzHugh--Nagumo (FHN) model \cite{fitzhughMathematicalModelsThreshold1955,fitzhughImpulsesPhysiologicalStates1961,nagumo-model}, a two-variable reduction of the Hodgkin--Huxley dynamics \cite{hh-old}, provides a minimal fast--slow description in which these noise-driven transitions can be studied systematically.

A further dynamical ingredient is introduced by finite signal-propagation and processing times. Delayed interactions enlarge the effective phase space of a dynamical system and can shift stability boundaries, generate or suppress oscillations, and produce multistability \cite{delays}. They can therefore strongly reorganize the parameter region in which noise-assisted response occurs. At the single-neuron level, an important source of delayed feedback is the autapse, a synaptic connection formed by a neuron onto itself \cite{autapse-proposition}. Autaptic connections have been reported in several cortical and subcortical regions \cite{autapse-hippocampus,autapse-neostriatum,autapse-substantia-nigra}, and delayed self-feedback models show that they can modify neuronal firing and stochastic resonance \cite{yilmaz2015455,yang2017}. From a dynamical-systems perspective, an autapse constitutes a local delayed feedback channel whose strength and timescale can be varied independently of the network topology, making it possible to distinguish local feedback effects from genuinely collective ones.

At the network level, most theoretical descriptions of noise-assisted neuronal dynamics have traditionally been formulated in terms of pairwise interactions. Many complex systems, however, contain irreducible interactions among groups of three or more units \cite{schneidman,ho-cortical,ho-cortical-2}. Such higher-order couplings can be represented by hypergraphs or simplicial complexes \cite{hypergraphs,simplex} and are known to alter collective states, synchronization thresholds, and stability properties in neuronal and other nonlinear networks \cite{ho-collective,ho-synchronization}.
The organization of interactions then becomes an additional control
parameter. In the additive diffusive formulation considered here, triadic hyperedges induce an effective weighted pairwise coupling, with weights determined by shared hyperedge membership. Consequently, hyperedge-based and ordinary pairwise networks can generate different collective responses even at the same nominal coupling strength, owing to differences in their effective connectivity and coupling weights. While recent studies have begun to investigate how higher-order interactions affect stochastic resonance \cite{Wang_2026,sr-2024}, their interplay with delayed local feedback and multilayer coupling remains largely unresolved.

An additional level of organization arises in multiplex systems, where several networks coexist and interact through interlayer connections. In the simplest duplex architecture, two network layers are coupled node to node, introducing a second collective interaction scale in addition to the intralayer dynamics \cite{multilayer-dynamics, multiplex-formal}. Such multilayer organization can generate dynamical regimes that have no direct single-layer counterpart \cite{multilayer-networks, duplex-chimera, triplex-chimera}. From the viewpoint of stochastic resonance, interlayer coupling raises a distinct question: rather than merely changing the response of a single network, it can transfer, redistribute, or equalize noise-assisted signal amplification between subsystems possessing different intrinsic resonance capacities. How this redistribution depends on interaction order, coupling strength, and transmission delay has not yet been systematically established.

Autaptic feedback, higher-order interactions, and multiplex organization therefore introduce three distinct levels of dynamical structure---local delayed feedback, intralayer interaction order, and interlayer coupling---whose effects on stochastic resonance need not be additive. Here we investigate their combined action in a delayed duplex network of excitable FHN neurons subject to weak periodic forcing and independent stochastic fluctuations. The model incorporates delayed autaptic self-feedback, delayed pairwise intralayer coupling, delayed triadic higher-order intralayer coupling, and delayed node-to-node interlayer coupling within a common framework. To disentangle the corresponding mechanisms, we construct the system hierarchically: starting from an isolated excitable neuron, we successively introduce autaptic feedback, pairwise and triadic interactions within a single layer, their combined action, and finally coupling between two network layers. Stochastic resonance is quantified through the spectral amplitude of the membrane-potential dynamics at the forcing frequency, together with its maximum over noise intensity and the associated optimal noise amplitude. Parameter regimes exhibiting repetitive deterministic spiking are identified separately so that the observed resonant response can be attributed to noise-assisted excitation.

This hierarchy reveals how the organization of interactions shapes the noise-assisted response. Among the autapse-free single-layer architectures, pairwise and triadic coupling attain comparable maximal resonance capacities, but differ in the range of nominal coupling strengths over which the enhancement persists; under the matched participation degrees used here, this difference largely reflects the doubled effective coupling associated with hyperedge-based connectivity.
Delayed autaptic feedback alone, or in networks containing exclusively pairwise or exclusively triadic interactions, predominantly shifts the noise intensity required for resonance without appreciably increasing the maximal response. Within the configurations investigated, a genuine autapse-induced enhancement emerges only when pairwise and triadic interactions coexist, demonstrating a non-additive interplay between local feedback and interaction order. In the duplex system, interlayer coupling primarily redistributes resonance between the two layers: weak-to-intermediate coupling can enhance an initially weakly resonating layer, but generally at the expense of the initially stronger one.
Interlayer coupling promotes response equalization, with the uncoupled resonance-capacity gap providing a useful descriptor of the observed redistribution. Increasing the interlayer delay predominantly suppresses the response within the sampled configurations and delay range.
Thus, the collective stochastic response is organized not by a single optimal coupling mechanism but by the competition among interaction order, resonance mismatch, and delayed information transfer.

The remainder of the paper is organized as follows. Section~\ref{sec:math_model_dynamics} introduces the stochastic delayed duplex FHN model and the hierarchy of interaction mechanisms. Section~\ref{sec:numerics} describes the numerical integration, simulation protocol, and spectral measure used to quantify stochastic resonance. Section~\ref{sec:results_discussion} presents the resonance properties of the isolated neuron, single-layer architectures, and duplex networks. Finally, Section~\ref{sec:summary_conclusions} summarizes the main conclusions.

\section{Mathematical model}
\label{sec:math_model_dynamics}
\subsection{Stochastic delayed duplex-network model}
\label{subsec:full_network_model}
We consider a delayed duplex network
\cite{yamakouControlCoherenceResonance2019,zheng2026optimized,nag2023interlayer}
composed of two layers of \(N\) identical FHN neurons
\cite{fitzhughMathematicalModelsThreshold1955,fitzhughImpulsesPhysiologicalStates1961}.
Neuron \(i=1,\ldots,N\) in layer \(\ell=1,2\) is described by the fast
membrane-potential variable \(v_i^{(\ell)}\) and the slow recovery
variable \(w_i^{(\ell)}\). The full stochastic system is:

\begin{equation}
\label{eq:full_model}
\left\{
\begin{aligned}
\frac{\mathrm{d}v_i^{(\ell)}}{\mathrm{d}t}
&=
v_i^{(\ell)}
\left(a-v_i^{(\ell)}\right)
\left(v_i^{(\ell)}-1\right)
-w_i^{(\ell)}
+\kappa_0^{(\ell)}
\left[
v_i^{(\ell)}(t-\tau_0^{(\ell)})
-v_i^{(\ell)}(t)
\right]
\\
&\quad
+\frac{\kappa_1^{(\ell)}}{N}
\sum_{j=1}^{N}
B_{ij}^{(\ell)}
\left[
v_j^{(\ell)}(t-\tau_1^{(\ell)})
-v_i^{(\ell)}(t)
\right]
\\
&\quad
+\frac{\kappa_2^{(\ell)}}{2N}
\sum_{j,k=1}^{N}
C_{ijk}^{(\ell)}
\left[
v_j^{(\ell)}(t-\tau_2^{(\ell)})
+v_k^{(\ell)}(t-\tau_2^{(\ell)})
-2v_i^{(\ell)}(t)
\right]
\\
&\quad
+\kappa_{\mathrm{12}}
\left[
v_i^{(3-\ell)}(t-\tau_{\mathrm{12}})
-v_i^{(\ell)}(t)
\right]
+A\cos(\Omega t)
+\sigma\,\frac{\mathrm{d}W_i^{(\ell)}}{\mathrm{d}t},
\\[6pt]
\frac{\mathrm{d}w_i^{(\ell)}}{\mathrm{d}t}
&=
\varepsilon
\left[
b\,v_i^{(\ell)}
-c\,w_i^{(\ell)}
\right],
\qquad
i=1,\ldots,N,\quad \ell=1,2 .
\end{aligned}
\right.
\end{equation}
The white-noise term in Eq.~\eqref{eq:full_model} is interpreted in the
It\^o sense,
\[
\mathrm{d}v_i^{(\ell)}
=
\mathcal{F}_i^{(\ell)}(t)\,\mathrm{d}t
+\sigma\,\mathrm{d}W_i^{(\ell)},
\]
where \(\mathcal{F}_i^{(\ell)}\) denotes the deterministic drift and the
\(W_i^{(\ell)}\) are mutually independent standard Wiener processes.
The required history functions for the delayed variables are specified
in Section~\ref{sec:history}.

The local FHN dynamics are fast--slow, with \(0<\varepsilon\ll1\).
Throughout this work, \(b=1\) and \(c=2\), while \(a\) controls excitability. In the isolated deterministic system without feedback, forcing, or noise, the origin is the unique stable equilibrium for
\(0<a<1+\sqrt{2}\). The values of \(a\) and \(\varepsilon\) used below lie in the excitable regime, and the periodic forcing is verified to be
subthreshold in Fig.~\ref{fig:subthreshold_regime}.

The four delayed interaction terms in Eq.~\eqref{eq:full_model}
represent autaptic, pairwise, triadic, and interlayer coupling,
respectively. Autaptic feedback has strength
\(\kappa_0^{(\ell)}\) and delay \(\tau_0^{(\ell)}\). Following the
signed delayed-feedback convention
\cite{autapse-negativecoupling1,autapse-negativecoupling2,
autapse-negativecoupling3,dynamics-electrical-autapse-hr},
\(\kappa_0^{(\ell)}<0\) and \(\kappa_0^{(\ell)}>0\) are termed
excitatory and inhibitory, respectively, according to their dynamical
effects; this phenomenological convention does not imply negative
physical conductances.

\begin{figure}[tbp]
\centering
\includegraphics[width=10cm]{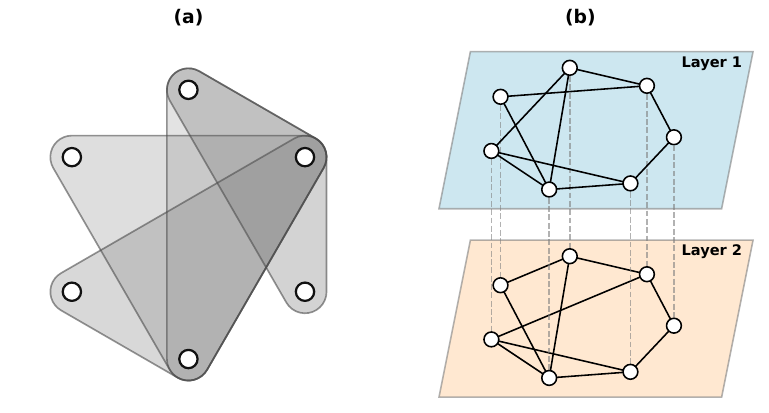}
\caption{Schematic of (a) triadic interactions based on 3-node hyperedges
and (b) the coupling structure of the duplex network.}
\label{fig:duplex_schematic}
\end{figure}

Pairwise interactions are encoded by the symmetric adjacency matrix
\(B^{(\ell)}\) of an undirected Watts--Strogatz network
\cite{small-world}, with \(B_{ii}^{(\ell)}=0\), coupling strength
\(\kappa_1^{(\ell)}\), and delay \(\tau_1^{(\ell)}\). Triadic
interactions \cite{hypergraphs,simplex} are encoded by a fully symmetric
tensor \(C^{(\ell)}\), with \(C_{ijk}^{(\ell)}=0\) whenever two indices
coincide. Each of the \(\binom{N}{3}\) unordered triplets is selected
independently with probability \(p_2\) according to an
Erd\H{o}s--R\'enyi-type three-body construction
\cite{erd-graph,er-graph2}, independently of the pairwise edges; hence,
a hyperedge need not form a \(2\)-simplex
[Fig.~\ref{fig:duplex_schematic}(a)]. Its coupling strength and delay
are \(\kappa_2^{(\ell)}\) and \(\tau_2^{(\ell)}\). The factor \(1/2\)
in Eq.~\eqref{eq:full_model} removes the double counting of
\((j,k)\) and \((k,j)\), while \(1/N\) provides the same global
network-size normalization as for the pairwise term.

The two layers are coupled node to node through delayed diffusive
interlayer coupling with strength \(\kappa_{\mathrm{12}}\) and delay
\(\tau_{\mathrm{12}}\), where \(v_i^{(3-\ell)}\) denotes the counterpart
of neuron \(i\) in the opposite layer
[Fig.~\ref{fig:duplex_schematic}(b)]. All coupling acts through the fast
variables. The system is driven by \(A\cos(\Omega t)\) and independent
noise of amplitude \(\sigma\); all neurons share the same intrinsic
parameters. A single-layer system is recovered by setting
\(\kappa_{\mathrm{12}}=0\) and restricting the dynamics to one layer.

\subsection{Initial conditions and history functions}
\label{sec:history}
Since Eq.~\eqref{eq:full_model} contains delays, the state history is
specified on \(t\in[-\tau_{\max},0]\), where
$
\tau_{\max}
=
\max_{\ell=1,2}
\left\{
\tau_0^{(\ell)},\tau_1^{(\ell)},\tau_2^{(\ell)},\tau_{12}
\right\}.
$
For the isolated-neuron simulations, we use the deterministic constant
history \(v(t)=0.2\), \(w(t)=0\). For configurations containing pairwise,
triadic, or interlayer coupling, each neuron instead has a constant
history
$
v_i^{(\ell)}(t)=v_{0,i}^{(\ell)},$  $w_i^{(\ell)}(t)=w_{0,i}^{(\ell)},$
with
$
v_{0,i}^{(\ell)}\sim\mathcal{N}(0.2,0.1^2)$, $w_{0,i}^{(\ell)}\sim\mathcal{N}(0,0.05^2),$
drawn independently across neurons and layers and also used as the
initial state at \(t=0\). Autaptic feedback is treated separately using an empty-loop
convention: its contribution is set to zero for
\(0\le t<\tau_0^{(\ell)}\) and activated for
\(t\ge\tau_0^{(\ell)}\), once simulated history is available.
The discarded transient \(T_0=3000\) is much longer than
the largest autaptic delay
(\(\tau_0^{(\ell)}\le 50\); see
Section~\ref{subsec:simulation_protocol}), excluding the
direct startup interval from the response calculation.
This timescale separation alone does not establish
independence from the initial history or feedback startup
convention.

\subsection{Model hierarchy and scope of the investigation}
\label{subsec:model_hierarchy}

Equation~\eqref{eq:full_model} provides a common framework for
distinguishing local autaptic feedback
\((\kappa_0^{(\ell)},\tau_0^{(\ell)})\),
intralayer connectivity through pairwise
\((\kappa_1^{(\ell)},\tau_1^{(\ell)})\)
and triadic
\((\kappa_2^{(\ell)},\tau_2^{(\ell)})\)
coupling, and interlayer communication
\((\kappa_{12},\tau_{12})\)
in noise-assisted signal response. The system is additionally controlled by the noise amplitude \(\sigma\), the periodic-forcing amplitude \(A\), and its angular
frequency \(\Omega\). Each interaction channel addresses a separate dynamical question. Delayed autaptic feedback couples a neuron to its own past voltage and isolates the influence of local memory without requiring interactions with other neurons. Pairwise intralayer coupling provides a reference for collective effects by coupling the voltage of each neuron to the delayed voltages of its neighbors. Triadic coupling instead organizes these contributions through three-node hyperedges. Because the triadic term in Eq.~\eqref{eq:full_model} is linear and additive in the two neighboring voltages, it can be represented equivalently as weighted pairwise diffusion, with weights determined by shared hyperedge membership. Thus, the distinction concerns the construction and weighting of connectivity, rather than an irreducible nonlinear three-body interaction.

Interlayer coupling introduces a further organizational level by connecting corresponding neurons in two layers. It allows us to examine how interactions between subsystems with different uncoupled resonance capacities enhance, suppress, or equalize their responses. All channels act through the voltage variable and have independently adjustable strengths and delays. We therefore introduce them progressively, from an isolated neuron to single-layer architectures and then duplex networks, setting the strengths of absent channels to zero. This hierarchy separates the effects of individual mechanisms and reveals how their combinations modify the response within the same local neuronal dynamics and forcing protocol.

Stochastic resonance is
identified through a nonmonotonic response to noise with a maximum at a
finite intermediate \(\sigma\); for every coupled configuration, the
deterministic dynamics at \(\sigma=0\) are examined separately, and
parameter regimes exhibiting repetitive deterministic spiking are
excluded from the classical SR analysis. These checks retain the periodic forcing and verify
subthreshold behavior for each coupled configuration
under the initialization specified in Section \ref{sec:history}.
They do not constitute a global exclusion of coexisting
attractors accessible from other history functions.

\section{Numerical methods}
\label{sec:numerics}

\subsection{Time discretization and stochastic integration}
\label{subsec:numerical_integration}
Equation~\eqref{eq:full_model} is integrated in the It\^{o} sense using the Euler--Maruyama method
\cite{ito,Maruyama1955,euler-maruyama}. With
\(t_n=n\Delta t\) and
\(\mathcal{F}_i^{(\ell)}(t_n)\) denoting the deterministic drift in the voltage equation, the discretization is
\begin{equation}
\label{eq:euler_maruyama}
\left\{
\begin{aligned}
v_{i,n+1}^{(\ell)}
&=
v_{i,n}^{(\ell)}
+\Delta t\,\mathcal{F}_i^{(\ell)}(t_n)
+\sigma\sqrt{\Delta t}\,\xi_{i,n}^{(\ell)},
\\[4pt]
w_{i,n+1}^{(\ell)}
&=
w_{i,n}^{(\ell)}
+\varepsilon\Delta t
\left[
b\,v_{i,n}^{(\ell)}
-c\,w_{i,n}^{(\ell)}
\right],
\end{aligned}
\right.
\end{equation}
where
\(\xi_{i,n}^{(\ell)}\sim\mathcal{N}(0,1)\) are mutually independent across neurons, layers, and time steps. The integration uses
\(\Delta t=0.01\). All delays are chosen as integer multiples of
\(\Delta t\), so delayed states are read directly from the stored history at \(n_\tau=\tau/\Delta t\) steps in the past without interpolation.

For each parameter configuration, the initial interval
\([0,T_0]\), with \(T_0=3000\), is discarded. Unless stated otherwise,
the common parameters are $\varepsilon=0.0015$, $a=0.153$, $b=1$, $c=2$, $A=0.01$, $\Omega=0.01$.
A spike is detected when \(v\) crosses
\(v_{\mathrm{th}}=0.25\) from below; this threshold is used only for
interspike intervals and for identifying deterministic repetitive
spiking and does not enter the spectral response measure \(Q\).

\begin{figure}[tbp]
\centering
\includegraphics[width=15cm]{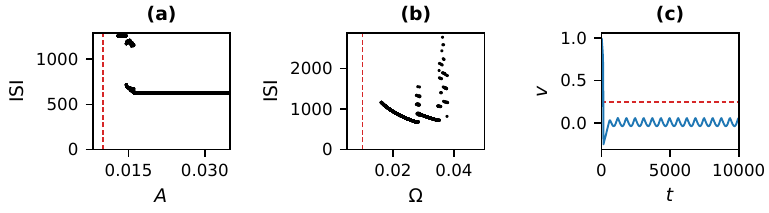}
\caption{Interspike-interval bifurcation diagrams for an isolated
neuron without autaptic feedback as functions of (a) the forcing
amplitude \(A\), with \(\Omega=0.01\), and (b) the angular frequency
\(\Omega\), with \(A=0.01\). Panel (c) shows the corresponding time
series of the fast variable \(v\) for \(A=0.01\) and \(\Omega=0.01\),
illustrating subthreshold oscillations in the absence of noise.}
\label{fig:subthreshold_regime}
\end{figure}

The forcing parameters are chosen to remain subthreshold for the
isolated deterministic neuron. Figure~\ref{fig:subthreshold_regime}(a)--(b)
shows the interspike-interval (ISI) bifurcation diagrams with respect to \(A\) and \(\Omega\),
while Fig.~\ref{fig:subthreshold_regime}(c) confirms that
\(A=0.01\) and \(\Omega=0.01\) produces only small-amplitude subthreshold
oscillations in the absence of noise.

\subsection{Quantification of stochastic resonance}
\label{subsec:sr_measure}
Stochastic resonance is quantified by the spectral amplitude of the
membrane-potential dynamics at the forcing frequency \(\Omega\), a
measure commonly used for single neurons \cite{hh-autapse-best-withq}
and neuronal networks \cite{Q-network,couplingstrength,sr-network-hh}.
After discarding the transient up to \(T_0\), we use an observation
window \(T_{\mathrm{obs}}=2\pi m/\Omega\), \(m\in\mathbb{N}\),
containing an integer number of forcing periods. For neuron \(i\) in
layer \(\ell\), the Fourier projections are
\begin{equation}
\label{eq:q_sin_cos}
\left\{
\begin{aligned}
Q_{\sin,i}^{(\ell)}
&=
\frac{2}{T_{\mathrm{obs}}}
\int_{T_0}^{T_0+T_{\mathrm{obs}}}
v_i^{(\ell)}(t)\sin(\Omega t)\,\mathrm{d}t,
\\
Q_{\cos,i}^{(\ell)}
&=
\frac{2}{T_{\mathrm{obs}}}
\int_{T_0}^{T_0+T_{\mathrm{obs}}}
v_i^{(\ell)}(t)\cos(\Omega t)\,\mathrm{d}t .
\end{aligned}
\right.
\end{equation}
The neuron-level amplitude and layer-averaged response are
\begin{equation}
\label{eq:q_neuron}
Q_i^{(\ell)}
=
\sqrt{
\left(Q_{\sin,i}^{(\ell)}\right)^2+
\left(Q_{\cos,i}^{(\ell)}\right)^2
},
\end{equation}
and
\begin{equation}
\label{eq:Q_av}
Q^{(\ell)}(\sigma)
=
\frac{1}{N}\sum_{i=1}^{N}Q_i^{(\ell)}(\sigma).
\end{equation}
For an isolated neuron, \(Q=Q_1\), whereas \(Q^{(1)}\) and \(Q^{(2)}\)
are evaluated separately in the duplex system. Since the modulus is
taken before averaging over neurons, \(Q^{(\ell)}\) measures the mean
response amplitude at the forcing frequency but not mutual neuronal
synchronization.

For each configuration, the noise amplitude is varied over
the investigated range $\mathcal{S}_\sigma = [10^{-4}, 0.15]$ with spacing \(\Delta\sigma\approx0.004\). The
resonance capacity and corresponding optimal noise amplitude are
defined by
\begin{equation}
\label{eq:qmax_sigmaopt}
Q_{\max}
=
\max_{\sigma\in\mathcal{S}_{\sigma}}Q(\sigma),
\qquad
\sigma_{\mathrm{opt}}
\in
\operatorname*{arg\,max}_{\sigma\in\mathcal{S}_{\sigma}}
Q(\sigma),
\end{equation}
so that \(\sigma_{\mathrm{opt}}\) is resolved to the noise-grid
spacing. A nonmonotonic \(Q(\sigma)\) with a maximum at a finite,
nonzero noise amplitude is taken as the numerical signature of SR.
Its interpretation is restricted to the excitable, subthreshold
regime, with parameter combinations exhibiting repetitive
deterministic spiking at \(\sigma=0\) excluded from the classical SR
analysis.

\subsection{Simulation protocol and statistical averaging}
\label{subsec:simulation_protocol}
Interaction mechanisms absent from a given configuration are removed by
setting the corresponding coupling strengths to zero. For the isolated
neuron, with or without autaptic feedback, simulations are performed up
to \(T_{\max}=10\,000\), and the response is averaged over
\(n_{\mathrm{traj}}=200\) independent noise realizations:
\begin{equation}
\label{eq:ensemble_average_q}
\langle Q(\sigma)\rangle
=
\frac{1}{n_{\mathrm{traj}}}
\sum_{r=1}^{n_{\mathrm{traj}}} Q_r(\sigma),
\end{equation}
where \(Q_r\) is the response of realization \(r\). Wherever this
ensemble average is used, \(Q\) denotes \(\langle Q\rangle\).

For single-layer and duplex-network simulations, the network size is
fixed at \(N=15\) and each parameter configuration is integrated up to
\(T_{\max}=100\,000\), providing long post-transient trajectories for
estimating the spectral response. The fixed size \(N=15\) reflects the computational cost
of combining long trajectories with scans over noise
amplitude, coupling strengths, and delays. Within the
available resources, long observation windows were
prioritized to reduce finite-observation-time variability
in the spectral response. No systematic comparison across
network sizes or ensemble of connectivity realizations
was performed; the reported network comparisons are
therefore conditional on the chosen size and connectivity.

For the network simulations, an additional ensemble average over long realizations at every point of the multidimensional parameter grid would be computationally prohibitive; we therefore use long post-transient trajectories and verify numerical stability by increasing the observation window and repeating representative simulations with independent Wiener-process realizations while keeping the pairwise and triadic network realizations fixed. The parameter scans cover delays \(\tau_0^{(\ell)}\in[0,50]\), \(\tau_1^{(\ell)},\tau_2^{(\ell)},\tau_{12}\in[0,10]\) and coupling strengths
\(\kappa_0^{(\ell)}\in[-1,1]\), \(\kappa_1^{(\ell)},\kappa_2^{(\ell)}\in[0,1]\), and \(\kappa_{12}\in\{0\}\cup[10^{-3},1]\), where the signed autaptic range includes excitatory and inhibitory feedback and \(\kappa_{12}=0\) defines the uncoupled duplex reference. Coupling strengths and delays are varied systematically while all other parameters are held fixed unless stated otherwise, thereby isolating their effects on \(Q_{\max}\) and \(\sigma_{\mathrm{opt}}\); each coupled configuration is additionally simulated at \(\sigma=0\) over the same parameter grid to identify deterministic repetitive-spiking regimes.

The parameter choices define a controlled comparison around a
deterministically subthreshold reference. The forcing parameters
\(A=0.01\) and \(\Omega=0.01\) are supported by the deterministic
scans in Fig.~\ref{fig:subthreshold_regime}, while the intrinsic parameters
\((a,\varepsilon)=(0.153,0.0015)\) are selected from Fig.~\ref{fig:single_neuron_sr_curves} to
provide a large isolated-neuron resonance response without
repetitive deterministic spiking. Holding these parameters fixed
in the network comparisons allows changes in the response to be
associated with the varied coupling channels.

The noise-amplitude interval
\(\sigma\in[10^{-4},0.15]\) resolves the resonance peaks and their
declining high-noise branches in the representative response
curves. The coupling-strength ranges cover weak interactions,
favorable intermediate regimes, and response suppression where observed. 

The delay intervals probe memory and transmission effects on
timescales shorter than the forcing period
\(T_{\mathrm{f}}=2\pi/\Omega\approx628\).
Specifically, the largest autaptic delay is approximately
\(0.08T_{\mathrm{f}}\), while the largest intralayer and interlayer
delays are approximately \(0.016T_{\mathrm{f}}\).
These ranges can nevertheless affect the fast voltage dynamics;
they do not explore delays comparable to or exceeding the forcing
period. All parameter values are expressed in model units and
are intended to characterize the investigated dynamical regimes,
rather than to calibrate a particular biological circuit.

\subsection{Network construction and participation-degree matching}
\label{subsec:density_matching}
Each layer is initialized as a regular ring lattice with degree \(k=4\),
and every eligible edge is rewired with probability \(\beta=0.25\).
Rewiring preserves the edge count, so \(\langle k\rangle=4\), although
individual node degrees vary.

The hyperedge probability \(p_2\) is fixed by matching expected node
participation across the two architectures. Neuron \(i\) belongs to
\(\binom{N-1}{2}\) possible triplets, so its expected second-order
participation degree is \(p_2\binom{N-1}{2}\); equating this to
\(\langle k\rangle\) gives
\begin{equation}
\label{eq:density_matching}
p_2
=
\frac{\langle k\rangle}{\binom{N-1}{2}}
=
\frac{2\langle k\rangle}{(N-1)(N-2)}
\approx 0.044
\quad\text{for } N=15,\ \langle k\rangle=4 .
\end{equation}

This construction matches the expected number of interaction
units incident on each neuron, with realized degrees fluctuating
across nodes. Both \(B^{(\ell)}\) and \(C^{(\ell)}\) are drawn
once and held fixed across a given parameter scan. Define
the pairwise degree and triadic participation degree as
\[
d_{1,i}^{(\ell)}=\sum_j B_{ij}^{(\ell)},
\qquad
d_{2,i}^{(\ell)}=\frac{1}{2}\sum_{j,k}C_{ijk}^{(\ell)}.
\]
The factor \(1/2\) removes double counting of each incident
hyperedge. Under Eq.~\eqref{eq:full_model}, the coefficients
multiplying the instantaneous negative self-voltage contributions
are therefore
\[
g_{1,i}^{(\ell)}
=\frac{\kappa_1^{(\ell)}d_{1,i}^{(\ell)}}{N},
\qquad
g_{2,i}^{(\ell)}
=\frac{2\kappa_2^{(\ell)}d_{2,i}^{(\ell)}}{N}.
\]
Thus, matching expected participation degrees does not equalize these coupling coefficients: at equal nominal strengths, the expected triadic coefficient is twice its pairwise
counterpart. 
Matching their expectations would instead require
\(\kappa_2^{(\ell)}=\kappa_1^{(\ell)}/2\), without making the
effective connectivity or nodewise coupling coefficients
identical. 
Thus, the common nominal interval $[0, 1]$ used for both scans covers twice the range of expected effective coupling in the triadic case; $\kappa_2^{(\ell)}=0.5$ corresponds to $\kappa_1^{(\ell)}=1$. This should be taken into account when comparing the pairwise and triadic response maps.

The shared factor \(1/N\) provides a common global
normalization, but neither degree normalization nor equality
of total coupling strength is imposed. Autaptic feedback,
as a single local channel, has no network-size normalization
and is assessed relative to the same architecture at
\(\kappa_0^{(\ell)}=0\). Accordingly, the comparisons describe
the specified architectures and feedback effects under the
adopted normalization; they do not establish an
interaction-order ranking at equal effective coupling strength
or under a fixed total coupling budget.

\section{Results and discussion}
\label{sec:results_discussion}

We now investigate how stochastic resonance changes as the system is
progressively extended from an isolated neuron to single-layer and
duplex-network architectures. The isolated neuron without feedback
serves as the reference configuration. We then examine the influence
of delayed autaptic feedback, pairwise and triadic intralayer
interactions, their combined action, and finally delayed interlayer
coupling. Throughout this section, resonance performance is assessed
using the resonance capacity \(Q_{\max}\) and the corresponding
optimal noise amplitude \(\sigma_{\mathrm{opt}}\).

These quantities distinguish the attainable response magnitude
from the noise amplitude required to attain it; a smaller
\(\sigma_{\mathrm{opt}}\) does not necessarily imply a larger
\(Q_{\max}\). Network comparisons use identical intrinsic
parameters and forcing, while autaptic effects are measured
relative to the corresponding architecture without feedback.
All maxima refer to the sampled noise grid and specified
coupling-parameter ranges, rather than unrestricted optima.
Comparisons between intralayer architectures are made at equal nominal coupling strength and matched expected participation degrees, which do not correspond to equal effective coupling (see Section~\ref{subsec:density_matching}).
\subsection{Isolated neuron without autaptic feedback}
\label{subsec:results_single_no_autapse}

\begin{figure}[tbp]
\centering
\includegraphics[width=10cm]{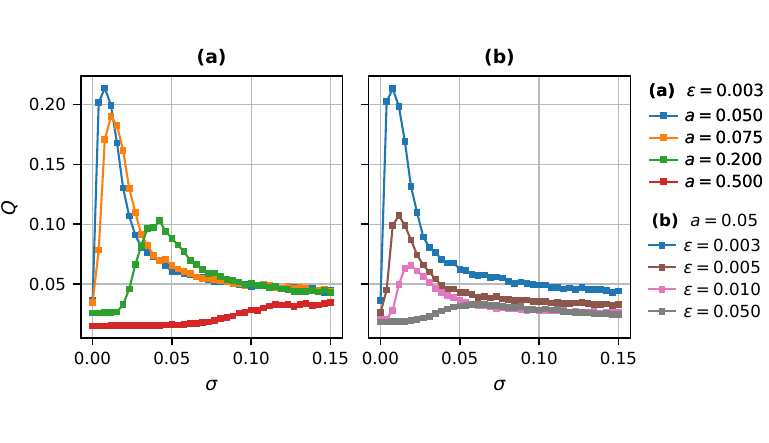}
\includegraphics[width=10cm]{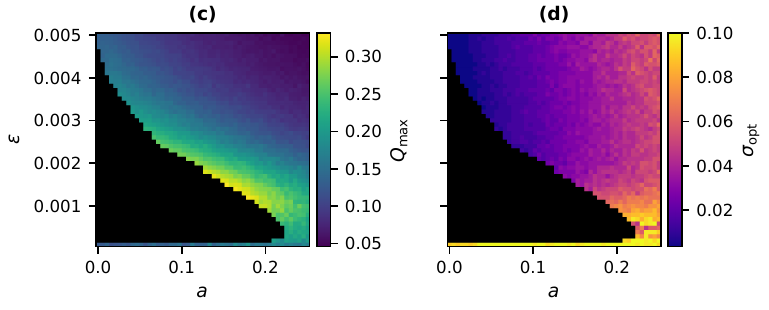}
\caption{Stochastic resonance in an isolated neuron without autaptic feedback. (a) Spectral response $Q(\sigma)$ for different values of $a$ at fixed $\varepsilon=0.003$, and (b) for different values of $\varepsilon$ at fixed $a=0.05$; each curve is averaged over $n_{\mathrm{traj}}=200$ independent noise realizations. (c) Resonance capacity $Q_{\max}$ and (d) optimal noise amplitude $\sigma_{\mathrm{opt}}$ as functions of $a$ and $\varepsilon$; the black region denotes repetitive deterministic spiking at $\sigma=0$.}
\label{fig:single_neuron_sr_curves}
\end{figure}

We first consider an isolated excitable neuron without autaptic
feedback, which serves as the reference configuration for the coupled
systems studied below. As shown in
Fig.~\ref{fig:single_neuron_sr_curves}, the spectral response
$Q(\sigma)$ exhibits the characteristic nonmonotonic dependence on the
noise amplitude: the response is weak for small $\sigma$, reaches a
maximum at an intermediate value $\sigma_{\mathrm{opt}}$, and decreases
again when strong noise destroys the coherence with the periodic
forcing.

At fixed $\varepsilon=0.003$, decreasing the excitability parameter
$a$ increases the resonance peak and shifts it toward smaller noise
amplitudes; see Fig.~\ref{fig:single_neuron_sr_curves}(a). Similarly,
at fixed $a=0.05$, decreasing $\varepsilon$ strengthens the resonance
and reduces the noise amplitude required to attain the maximum; see
Fig.~\ref{fig:single_neuron_sr_curves}(b). Thus, neurons closer to the
excitation threshold and with stronger timescale separation exhibit a
larger and more noise-efficient response.

The joint dependence on the intrinsic parameters is summarized in
Figs.~\ref{fig:single_neuron_sr_curves}(c) and \ref{fig:single_neuron_sr_curves}(d). The largest values of
$Q_{\max}$ occur close to the boundary of deterministic excitation,
whereas configurations farther inside the quiescent regime generally
require stronger noise and exhibit a weaker maximum response. The black region corresponds to parameter combinations that already produce repetitive spiking at $\sigma=0$ and is therefore excluded from the classical stochastic-resonance analysis.

The deterministically subthreshold configuration
$(a,\varepsilon)=(0.153,0.0015)$ yields
$Q_{\max}\approx0.33$ and is adopted for the subsequent analysis. This
value serves as the isolated-neuron reference level for assessing
whether autaptic, intralayer, and interlayer coupling enhance or
suppress stochastic resonance.

\subsection{Isolated neuron with delayed autaptic feedback}
\label{subsec:results_single_autapse}
\begin{figure}[tbp]
\centering
\includegraphics[width=10cm]{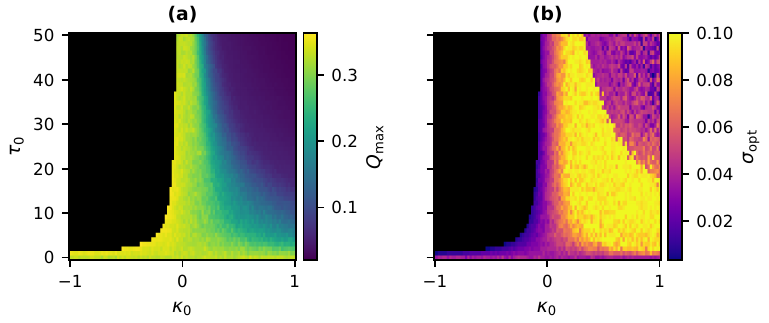}
\includegraphics[width=10cm]{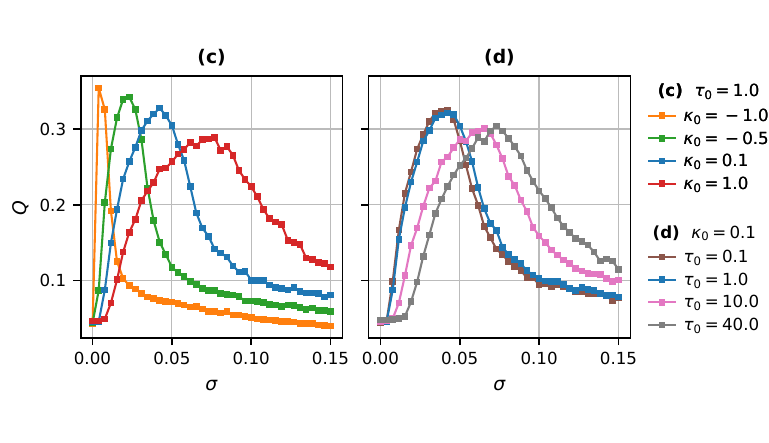}
\caption{Stochastic resonance in an isolated neuron with delayed autaptic feedback, with intrinsic parameters $(a,\varepsilon)=(0.153,0.0015)$. (a) Resonance capacity $Q_{\max}$ and (b) optimal noise amplitude
$\sigma_{\mathrm{opt}}$ as functions of the autaptic coupling strength $\kappa_0$ and delay $\tau_0$; the black region denotes repetitive deterministic spiking at $\sigma=0$.
(c) Representative spectral-response curves $Q(\sigma)$ for different coupling strengths $\kappa_0$ at fixed $\tau_0=1$, and (d) for different delays $\tau_0$ at fixed $\kappa_0=0.1$. Each curve is averaged over $n_{\mathrm{traj}}=200$ independent noise realizations.}
\label{fig:single_neuron_autapse_maps}
\end{figure}

We next examine whether delayed autaptic feedback can improve the SR
response of the isolated neuron with
$(a,\varepsilon)=(0.153,0.0015)$. Figure~\ref{fig:single_neuron_autapse_maps}(a)--(b)
shows that the largest values of $Q_{\max}$ occur close to the boundary
of deterministic excitation. Weak excitatory feedback
($\kappa_0<0$) with a short delay produces only a marginal increase
above the autapse-free reference value $Q_{\max}\approx0.33$, but substantially reduces the corresponding optimal noise amplitude.
The parameter region supporting this improvement is narrow, and
stronger excitatory feedback or larger delays can induce repetitive
spiking already at $\sigma=0$; these configurations are excluded from the classical SR analysis.

The representative $Q(\sigma)$ curves in
Fig.~\ref{fig:single_neuron_autapse_maps}(c)--(d) confirm that increasing
excitatory feedback shifts the resonance peak toward weaker noise.
In contrast, inhibitory feedback ($\kappa_0>0$) generally reduces
$Q_{\max}$. For weak inhibitory coupling, increasing $\tau_0$ mainly
shifts the resonance peak toward larger noise amplitudes, whereas
strong inhibitory feedback combined with a large delay markedly
suppresses the response. Away from the deterministic-excitation
boundary, $\sigma_{\mathrm{opt}}$ remains approximately $0.09$, but
the corresponding resonance capacity is considerably lower.

Thus, delayed autaptic feedback does not substantially increase the
maximum resonance capacity of the optimized isolated neuron. Its main
effect is to shift the noise amplitude required for resonance, while
strong feedback may either induce deterministic spiking or suppress
SR. The suboptimal intrinsic-neuron configuration exhibits the same
qualitative dependence on $(\kappa_0,\tau_0)$ but a substantially
smaller resonance capacity and is therefore omitted here.

\subsection{Single-layer networks without autaptic feedback}
\label{subsec:results_intralayer_no_autapse}

This subsection compares the effects of pairwise and triadic higher-order interactions before delayed autaptic feedback is introduced. It is useful to separate the three architectures below because they exhibit qualitatively different dependencies on coupling strength and delay.

\subsubsection{Networks with pairwise interactions}
\label{subsubsec:results_pairwise}

\begin{figure}
\centering
\includegraphics[width=10cm]{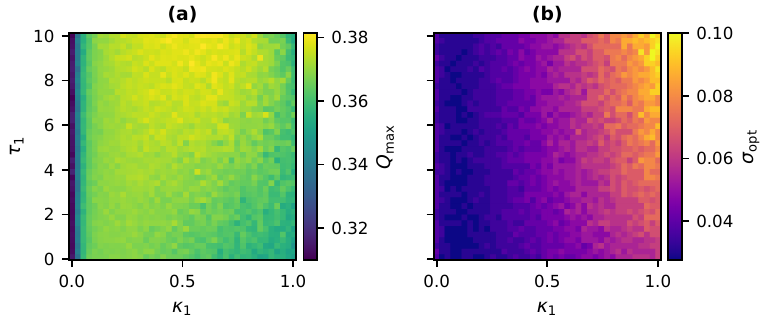}
\includegraphics[width=10cm]{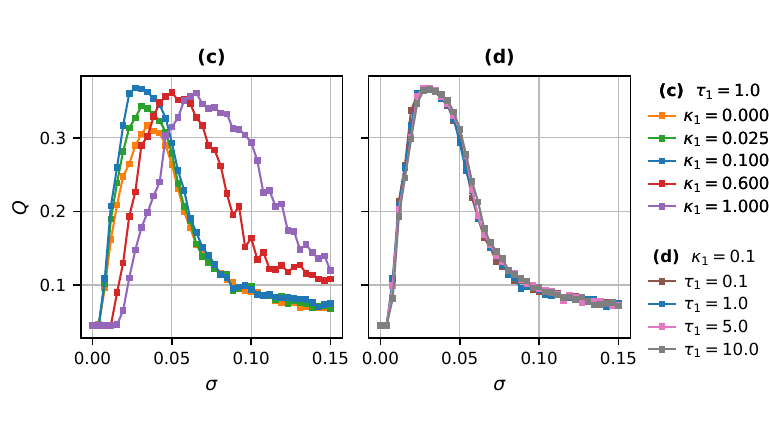}
\caption{Stochastic resonance in a single-layer network with pairwise interactions.
(a) Resonance capacity $Q_{\max}$ and (b) optimal noise amplitude $\sigma_{\mathrm{opt}}$ as functions of the coupling strength $\kappa_1$ and delay $\tau_1$. (c) Spectral response $Q(\sigma)$ for different coupling strengths $\kappa_1$
at fixed delay $\tau_1=1$, and (d) for different delays $\tau_1$ at fixed coupling strength $\kappa_1=0.1$.}
\label{fig:pairwise_parameter_maps}
\end{figure}

We next consider a single-layer network with pairwise interactions and
without autaptic feedback. As shown in
Fig.~\ref{fig:pairwise_parameter_maps}(a)--(b), pairwise coupling enhances the
resonance capacity above the isolated-neuron reference
$Q_{\max}\approx0.33$. For weak coupling,
$\kappa_1<0.05$, the network response approaches the isolated-neuron
baseline, whereas $Q_{\max}$ increases with coupling and reaches its
largest value, approximately $0.38$, near
$(\tau_1,\kappa_1)=(10,0.5)$. Stronger coupling subsequently reduces
the maximum response, revealing an optimal intermediate
collective-coupling regime.

The representative curves in
Fig.~\ref{fig:pairwise_parameter_maps}(c)--(d) further illustrate these trends.
At $\tau_1=1$, varying $\kappa_1$ changes both the height and
position of the resonance peak, with intermediate coupling producing
the strongest response. By contrast, for weak coupling
$\kappa_1=0.1$, the curves obtained for different delays nearly
coincide, indicating that $\tau_1$ has little influence in this regime. Overall, neither very weak nor excessively strong coupling is optimal; pairwise interactions enhance SR most effectively when
the coupling strength is tuned to an intermediate value.

\subsubsection{Triadic higher-order interactions with and without pairwise coupling}
\label{subsubsec:results_triadic_pairwise}

\begin{figure}[tbp]
\centering
\includegraphics[width=10cm]{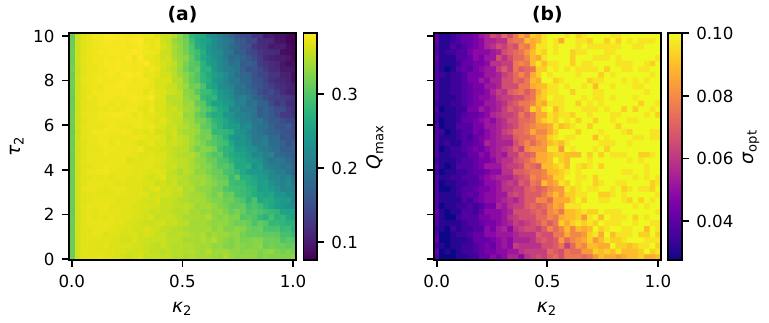}
\includegraphics[width=10cm]{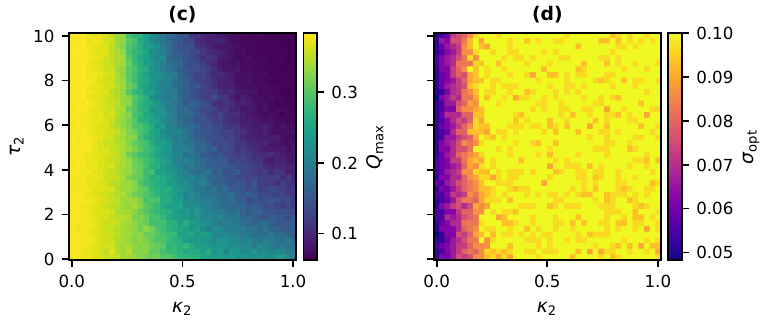}
\caption{Stochastic-resonance characteristics of single-layer networks
with triadic higher-order interactions as functions of the coupling
strength \(\kappa_2\) and delay \(\tau_2\). Panels (a)--(b) show
\(Q_{\max}\) and \(\sigma_{\mathrm{opt}}\), respectively, without
pairwise interactions; panels (c)--(d) show the corresponding quantities
with fixed pairwise coupling \((\tau_1,\kappa_1)=(10,0.5)\). Note that, under the matched participation degrees of Section~\ref{subsec:density_matching}, $\kappa_2$ corresponds in expectation to the same effective coupling as a pairwise strength $\kappa_1=2\kappa_2$.}
\label{fig:triadic_pairwise_parameter_maps}
\end{figure}

We next examine triadic higher-order interactions, first alone and then
in combination with the optimized pairwise configuration. As shown in
Fig.~\ref{fig:triadic_pairwise_parameter_maps}(a), weak triadic coupling enhances the resonance capacity above the isolated-neuron reference, while the enhancement at intermediate \(\kappa_2\) becomes delay dependent. Beyond the nearly uncoupled regime,
\(Q_{\max}\) decreases with increasing \(\kappa_2\) and is strongly
suppressed when both coupling strength and delay are large, in contrast
to the pairwise network whose optimum occurs at intermediate coupling.
Concurrently, \(\sigma_{\mathrm{opt}}\) increases with \(\kappa_2\),
especially for \(\kappa_2\gtrsim0.5\)
[Fig.~\ref{fig:triadic_pairwise_parameter_maps}(b)]. Thus, within the triadic architecture, the largest resonance capacity and the lowest noise requirement both occur at weak coupling.

When triadic interactions are added to the optimized pairwise network,
\((\tau_1,\kappa_1)=(10,0.5)\), the region of large \(Q_{\max}\) in the
\((\tau_2,\kappa_2)\) plane becomes narrower
[Fig.~\ref{fig:triadic_pairwise_parameter_maps}(c)]. The enhanced
pairwise response is nevertheless preserved as \(\kappa_2\to0\),
whereas increasing \(\kappa_2\) progressively suppresses \(Q_{\max}\)
and shifts the optimum toward larger noise amplitudes
[Fig.~\ref{fig:triadic_pairwise_parameter_maps}(d)]. Hence, the
pairwise-enhanced SR state is robust to weak higher-order perturbations
but deteriorates under strong triadic coupling.

\subsubsection{Comparison of autapse-free single-layer architectures}
\label{subsubsec:results_architecture_comparison}
Both autapse-free architectures enhance the response above the isolated-neuron reference $Q_{\max}\approx0.33$. Pairwise coupling attains $Q_{\max}\approx0.38$ at $(\tau_1,\kappa_1)=(10,0.5)$ and the triadic architecture attains a comparable capacity in its weak-coupling regime; the difference between the two largest observed values is not resolved beyond the residual variability of the response measure estimated in Section~\ref{subsubsec:results_triadic_autapse}. The architectures differ instead in the range of nominal coupling strengths over which the enhancement persists. The pairwise enhancement is maintained across the investigated range of $\kappa_1$ beyond the weakest couplings, whereas the triadic enhancement is confined to weak and intermediate $\kappa_2$ and is progressively suppressed at larger $\kappa_2$, most strongly in combination with large delays. Correspondingly, $\sigma_{\mathrm{opt}}$ increases only gradually with $\kappa_1$ but more rapidly with $\kappa_2$.

This comparison is made at equal nominal coupling strength and matched expected participation degrees, which do not correspond to equal effective coupling (Section~\ref{subsec:density_matching}). Since the expected triadic self-voltage coefficient is twice its pairwise counterpart, the scanned interval $\kappa_2 \in [0,1]$ covers twice the range of expected effective coupling covered by $\kappa_1 \in [0,1]$, with $\kappa_2 = 0.5$ corresponding to $\kappa_1 = 1$. The suppression of the triadic response occurs predominantly at $\kappa_2 \gtrsim 0.5,$ that is, largely at effective coupling strengths beyond those sampled in the pairwise architecture, while both architectures are enhanced within the effective range sampled by both. The more restricted range of triadic enhancement in nominal units is therefore to a large extent attributable to this normalization, and the present scans do not determine whether the pairwise architecture would exhibit comparable suppression at equally strong effective coupling.

The combined architecture preserves the pairwise enhancement under weak triadic coupling, whereas strong triadic coupling reduces $Q_{\max}$ and shifts $\sigma_{\mathrm{opt}}$ toward larger noise amplitudes.
A qualitative interpretation follows from the competing
effects of diffusive coupling. Delayed input from an excited
neighbor can promote excitation of a resting neuron, whereas
the same coupling restrains a neuron whose voltage exceeds
the delayed neighboring voltages. Intermediate pairwise
coupling can therefore favor the recruitment of noise-triggered
excursions during favorable forcing phases, consistent
with the increased \(Q_{\max}\) in Fig.~\ref{fig:pairwise_parameter_maps}.

For the additive triadic term, the corresponding recruitment
and restraint act through the effective weighted pairwise
connectivity induced by hyperedge membership.
The low \(\sigma_{\mathrm{opt}}\) at weak triadic coupling
is consistent with coupling-assisted excitation, while
the decline of \(Q_{\max}\) and increase of
\(\sigma_{\mathrm{opt}}\) at stronger coupling are consistent
with increased restraint of excursions. This interpretation
is qualitative: the response maps do not separately identify
the contributions of coupling weights, topology, and delay,
or establish an interaction-order-specific mechanism.

\subsection{Single-layer networks with delayed autaptic feedback}
\label{subsec:results_intralayer_autapse}

We next determine whether delayed autaptic feedback can further
improve the resonance generated by intralayer interactions.

\subsubsection{Pairwise networks with autaptic feedback}
\label{subsubsec:results_pairwise_autapse}

\begin{figure}[tbp]
\centering
\includegraphics[width=10cm]{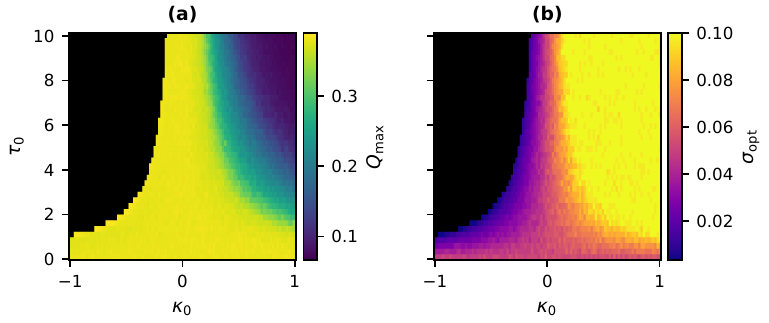}
\includegraphics[width=10cm]{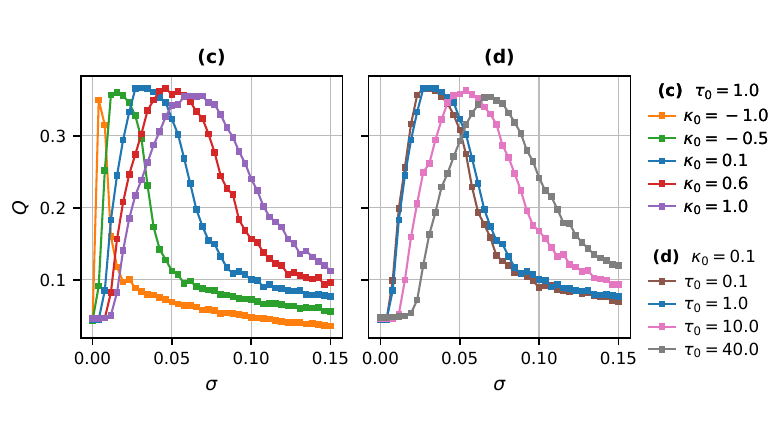}
\caption{Effect of delayed autaptic feedback on a pairwise-coupled
single-layer network with fixed interaction parameters
\((\tau_1,\kappa_1)=(1,0.1)\).
(a) Resonance capacity \(Q_{\max}\) and (b) optimal noise amplitude
\(\sigma_{\mathrm{opt}}\) as functions of the autaptic coupling
strength \(\kappa_0\) and delay \(\tau_0\); the black region denotes
repetitive deterministic spiking at \(\sigma=0\).
(c) Spectral response \(Q(\sigma)\) for different autaptic coupling
strengths \(\kappa_0\) at fixed \(\tau_0=1\), and (d) for different
autaptic delays \(\tau_0\) at fixed \(\kappa_0=0.1\).}
\label{fig:pairwise_autapse_maps}
\end{figure}

We consider a pairwise network with fixed intralayer parameters
\((\tau_1,\kappa_1)=(1,0.1)\) and vary the autaptic strength
\(\kappa_0\) and delay \(\tau_0\). As shown in
Fig.~\ref{fig:pairwise_autapse_maps}(a)--(b), autaptic feedback does
not appreciably increase the resonance capacity above that of the
corresponding autapse-free network. Excitatory feedback with short
delay preserves a large \(Q_{\max}\) while shifting
\(\sigma_{\mathrm{opt}}\) toward weaker noise, whereas stronger
excitatory feedback or larger delays can induce repetitive
deterministic spiking. Inhibitory feedback instead progressively
reduces \(Q_{\max}\) and shifts the optimum toward larger noise
amplitudes, with the strongest suppression occurring for large
\(\kappa_0\) and \(\tau_0\).

The representative curves in
Fig.~\ref{fig:pairwise_autapse_maps}(c)--(d) confirm that the dominant
effect of the autapse is a displacement of the resonance peak along
the noise axis. At fixed \(\tau_0=1\), increasing \(\kappa_0\) from
excitatory to inhibitory values shifts the optimum toward larger
\(\sigma\) while changing the peak height only moderately over a broad range; increasing \(\tau_0\) at fixed \(\kappa_0=0.1\) produces a
similar shift. Thus, delayed autaptic feedback primarily tunes the
noise scale for SR rather than enhancing its maximal response, while
strong inhibitory feedback and long delays suppress the resonance.

\subsubsection{Triadic networks with autaptic feedback}
\label{subsubsec:results_triadic_autapse}

\begin{figure}[tbp]
\centering
\includegraphics[width=10cm]{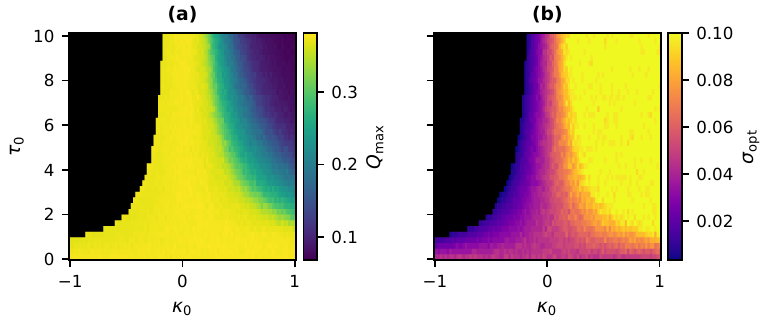}
\includegraphics[width=10cm]{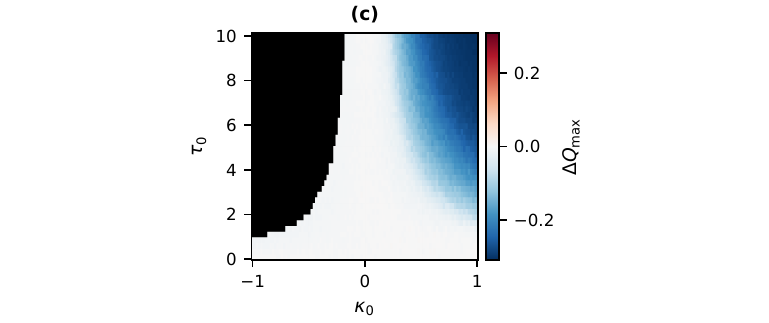}
\caption{Effect of delayed autaptic feedback on a single-layer network
with fixed triadic interactions
$(\tau_2,\kappa_2)=(8,0.2)$:
(a) resonance capacity $Q_{\max}$,
(b) optimal noise amplitude $\sigma_{\mathrm{opt}}$, and
(c) change in resonance capacity
$\Delta Q_{\max}
=Q_{\max}^{\mathrm{aut}}-Q_{\max}^{\mathrm{no\,aut}}$
relative to the corresponding autapse-free triadic network. The black
region denotes repetitive deterministic spiking at $\sigma=0$.}
\label{fig:triadic_autapse_maps}
\end{figure}

We next introduce delayed autaptic feedback into a network with fixed
triadic interactions at $(\tau_2,\kappa_2)=(8,0.2)$. As shown in
Fig.~\ref{fig:triadic_autapse_maps}(a), varying the autaptic coupling
strength $\kappa_0$ and delay $\tau_0$ does not produce a resolvable increase in the resonance capacity beyond that of the corresponding autapse-free triadic network. Excitatory feedback close to the deterministic-excitation boundary may preserve a relatively large response, but stronger feedback or increasing delay can induce repetitive deterministic
spiking. As for the pairwise interactions, inhibitory feedback generally reduces $Q_{\max}$.

The optimal-noise map in
Fig.~\ref{fig:triadic_autapse_maps}(b) shows that the autapse can shift
the noise amplitude required for resonance, although this shift is not
accompanied by an improvement in the attainable maximum response. This
is confirmed by the difference
\begin{equation}\label{DQ}
\Delta Q_{\max}
=
Q_{\max}^{\mathrm{aut}}
-
Q_{\max}^{\mathrm{no\,aut}}.    
\end{equation}
As shown in Fig.~\ref{fig:triadic_autapse_maps}(c), no numerically significant positive enhancement of the resonance capacity is resolved within the deterministically subthreshold parameter region. The largest positive excursion, \(\Delta Q_{\max}\approx 0.003\), is smaller than the residual variation of \(Q_{\max}\) observed near \(\kappa_0=0\) (approximately \(0.005\)), where the autapse is inactive and \(Q_{\max}\) should therefore be independent of \(\tau_0\). We therefore regard this small positive excursion as being within the numerical variability of the response measure rather than as evidence of a genuine autapse-induced enhancement. Thus, as in the pairwise network, delayed autaptic feedback mainly modifies the noise amplitude required for resonance and does not produce a resolvable increase in the resonance capacity of a network containing only triadic interactions.

\subsubsection{Combined pairwise--triadic networks with autaptic feedback}
\label{subsubsec:results_combined_autapse}

We finally introduce delayed autaptic feedback into the combined
pairwise--triadic network with fixed interaction parameters
$(\tau_1,\kappa_1)=(10,0.5)$ and
$(\tau_2,\kappa_2)=(8,0.2)$. In contrast to the pairwise-only and
triadic-only architectures, this configuration exhibits a clear
autapse-induced enhancement of the resonance capacity.

\begin{figure}[tbp]
\centering
\includegraphics[width=10cm]{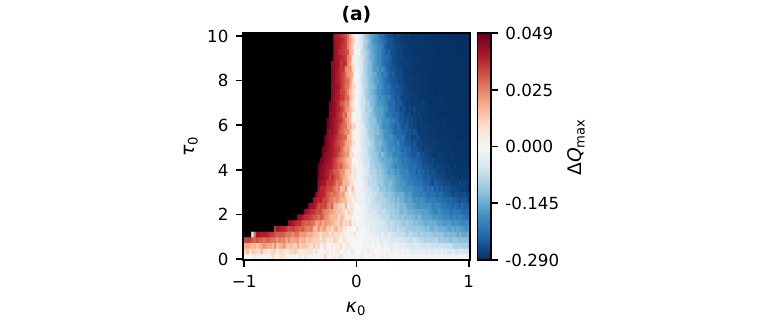}
\includegraphics[width=10cm]{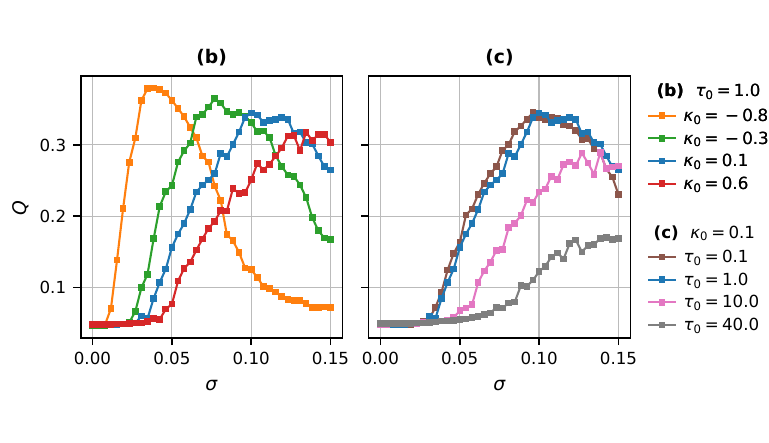}
\caption{Effect of delayed autaptic feedback on a single-layer network with fixed pairwise interactions $(\tau_1,\kappa_1)=(10,0.5)$ and triadic interactions $(\tau_2,\kappa_2)=(8,0.2)$. (a) Change in resonance capacity $\Delta Q_{\max}=Q_{\max}^{\mathrm{aut}}-Q_{\max}^{\mathrm{no\,aut}}$ as a function of the autaptic coupling strength $\kappa_0$ and delay $\tau_0$; positive values indicate an enhancement relative to the corresponding
autapse-free network, and the black region denotes repetitive deterministic spiking at $\sigma=0$. (b) Spectral response $Q(\sigma)$ for different autaptic coupling strengths $\kappa_0$ at fixed $\tau_0=1$, and (c) for different autaptic delays $\tau_0$ at fixed $\kappa_0=0.1$.}
\label{fig:combined_autapse_results}
\end{figure}

To quantify this effect, we use $\Delta Q_{\max}$ defined in Eq.~\eqref{DQ}.
As shown in Fig.~\ref{fig:combined_autapse_results}(a), excitatory
autaptic feedback ($\kappa_0<0$) produces a positive
$\Delta Q_{\max}$ in a parameter region close to the boundary of
deterministic excitation. The maximum increase is approximately
$\Delta Q_{\max}\approx0.048$, corresponding to a gain of about
$15\%$ relative to the same pairwise--triadic network without autaptic feedback. Inhibitory feedback ($\kappa_0>0$), by contrast, does not produce a comparable enhancement and generally reduces the
resonance capacity.

The representative curves in
Fig.~\ref{fig:combined_autapse_results}(b) confirm that, for the short
delay $\tau_0=1$, increasing the excitatory autaptic strength raises
the SR peak and shifts it toward smaller noise amplitudes. The
strongest response is obtained for the largest deterministically
subthreshold excitatory coupling considered, whereas inhibitory
feedback lowers the peak and shifts the optimum toward stronger
noise.

Figure~\ref{fig:combined_autapse_results}(c) illustrates the influence
of the autaptic delay at fixed weak inhibitory coupling
$\kappa_0=0.1$. Short delays produce comparable resonance curves,
whereas increasing $\tau_0$ progressively lowers the peak and shifts
it toward larger noise amplitudes. In particular, the response is strongly suppressed for $\tau_0=40$.

These results show that within the configurations investigated, a clear autapse-induced enhancement emerges only when pairwise and triadic interactions coexist. The enhancement is therefore not a generic effect of self-feedback, but indicates a non-additive interplay among autaptic, pairwise, and higher-order coupling mechanisms. Nevertheless, this benefit is restricted to excitatory feedback with sufficiently short delays, while long delays and inhibitory feedback are detrimental to SR.

\subsection{Duplex networks}
\label{subsec:results_duplex}

We finally investigate whether interlayer coupling can improve the SR
response of a weakly resonating layer by coupling it to a layer with a
larger uncoupled resonance capacity. For each duplex configuration,
the intralayer parameters are chosen such that, in the uncoupled limit,
layer~1 has the larger resonance capacity,
$Q_{\max}^{(1)}>Q_{\max}^{(2)}$, and serves as the reference layer,
whereas layer~2 exhibits a weaker response. We then vary the interlayer
coupling strength $\kappa_{\mathrm{12}}$ and delay
$\tau_{\mathrm{12}}$ to determine how the resonance characteristics
are redistributed between the two layers.

Table~\ref{tab:duplex_configurations} summarizes the six duplex
configurations considered in this study. Configurations~1--4 comprise
all ordered pairings of layers containing exclusively pairwise or
triadic interactions. Configuration~5 combines pairwise and triadic
interactions in both layers, whereas configuration~6 additionally
includes delayed autaptic feedback. In every configuration, the
intralayer parameters are chosen such that layer~1 has the larger
uncoupled resonance capacity.

\begin{table}
\centering
\caption{Duplex-network configurations considered in the numerical
simulations. The parameters $\kappa_0$, $\kappa_1$, and $\kappa_2$
denote the autaptic, pairwise, and triadic coupling strengths,
respectively, while $\tau_0$, $\tau_1$, and $\tau_2$ denote the
corresponding delays. A dash indicates that the corresponding
interaction mechanism is absent.}
\label{tab:duplex_configurations}
\tiny
\setlength{\tabcolsep}{4pt}
\renewcommand{\arraystretch}{1.2}
\resizebox{\textwidth}{!}{
\begin{tabular}{c cccccc cccccc}
\toprule
\multirow{2}{*}{\textbf{Config.}}
&
\multicolumn{6}{c}{\textbf{Layer 1}}
&
\multicolumn{6}{c}{\textbf{Layer 2}}
\\
\cmidrule(lr){2-7}
\cmidrule(lr){8-13}
&
$\kappa_0^{(1)}$
&
$\tau_0^{(1)}$
&
$\kappa_1^{(1)}$
&
$\tau_1^{(1)}$
&
$\kappa_2^{(1)}$
&
$\tau_2^{(1)}$
&
$\kappa_0^{(2)}$
&
$\tau_0^{(2)}$
&
$\kappa_1^{(2)}$
&
$\tau_1^{(2)}$
&
$\kappa_2^{(2)}$
&
$\tau_2^{(2)}$
\\
\midrule

1
&
$-$ & $-$ & 0.5 & 10.0 & $-$ & $-$
&
$-$ & $-$ & 0.01 & 1.0 & $-$ & $-$
\\

2
&
$-$ & $-$ & 0.5 & 10.0 & $-$ & $-$
&
$-$ & $-$ & $-$ & $-$ & 1.0 & 10.0
\\

3
&
$-$ & $-$ & $-$ & $-$ & 0.2 & 8.0
&
$-$ & $-$ & 0.01 & 1.0 & $-$ & $-$
\\

4
&
$-$ & $-$ & $-$ & $-$ & 0.2 & 8.0
&
$-$ & $-$ & $-$ & $-$ & 1.0 & 10.0
\\

5
&
$-$ & $-$ & 0.5 & 10.0 & 0.1 & 1.0
&
$-$ & $-$ & 0.5 & 10.0 & 1.0 & 10.0
\\

6
&
$-0.33$ & 4.0 & 0.5 & 10.0 & 0.2 & 8.0
&
1.0 & 10.0 & 0.5 & 10.0 & 0.2 & 8.0
\\

\bottomrule
\end{tabular}
}
\end{table}

In the sequel, the effects of interlayer coupling strength and delay are compared across all six configurations.

\subsubsection{Homogeneous and heterogeneous duplex architectures}
\label{subsubsec:results_duplex_architectures}

\begin{figure}[tbp]
\centering
\includegraphics[]{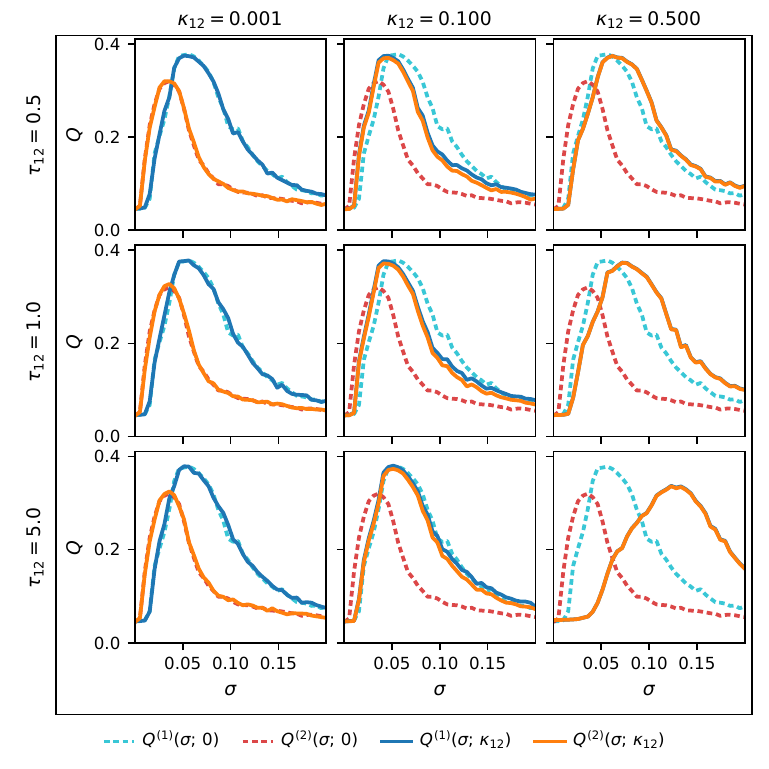}
\caption{Representative SR curves for the homogeneous
pairwise--pairwise duplex of configuration~1 in
Table~\ref{tab:duplex_configurations} at selected interlayer coupling
strengths and delays. Columns correspond to
\(\kappa_{\mathrm{12}}=0.001\), \(0.1\), and \(0.5\), and rows to
\(\tau_{\mathrm{12}}=0.5\), \(1.0\), and \(5.0\). Dashed and solid
curves denote the uncoupled responses
[\(Q^{(1)}(\sigma;0)\), \(Q^{(2)}(\sigma;0)\)] and coupled responses
[\(Q^{(1)}(\sigma;\kappa_{12})\), \(Q^{(2)}(\sigma;\kappa_{12})\)],
respectively.}
\label{fig:duplex_homogeneous}
\end{figure}

\begin{figure}[tbp]
\centering
\includegraphics[]{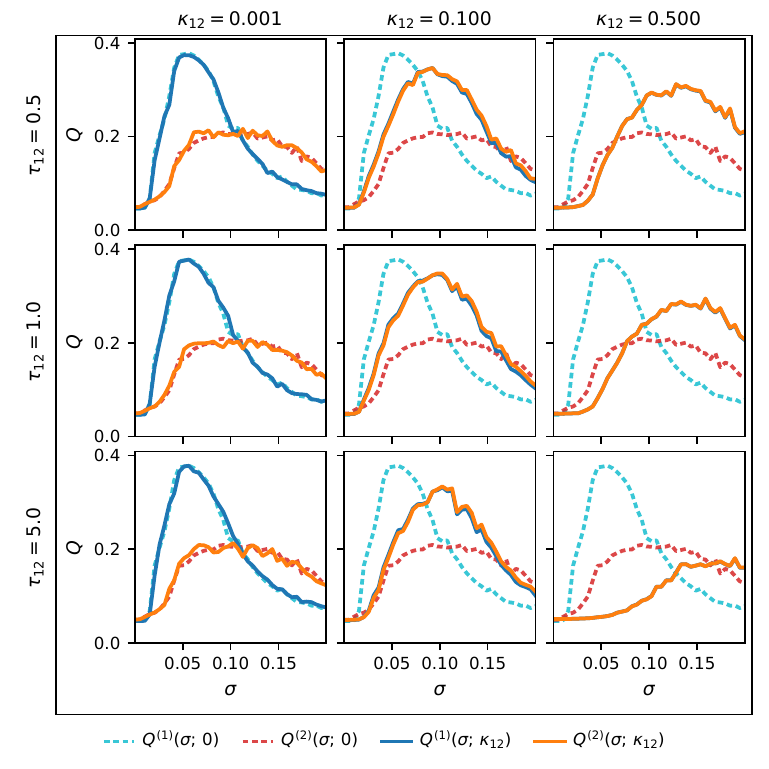}
\caption{SR curves for the heterogeneous pairwise--triadic duplex of
configuration~2 in Table~\ref{tab:duplex_configurations} at selected
interlayer coupling strengths and delays. Columns correspond to
\(\kappa_{\mathrm{12}}=0.001\), \(0.1\), and \(0.5\), and rows to
\(\tau_{\mathrm{12}}=0.5\), \(1.0\), and \(5.0\). Dashed and solid
curves denote the uncoupled responses
[\(Q^{(1)}(\sigma;0)\), \(Q^{(2)}(\sigma;0)\)] and coupled responses
[\(Q^{(1)}(\sigma;\kappa_{12})\), \(Q^{(2)}(\sigma;\kappa_{12})\)],
respectively.}
\label{fig:duplex_heterogeneous}
\end{figure}

We compare a homogeneous duplex, in which both layers have the same
interaction architecture, with a heterogeneous duplex composed of
layers with different interaction orders. In
Figs.~\ref{fig:duplex_homogeneous} and
\ref{fig:duplex_heterogeneous}, dashed and solid curves denote the
uncoupled and coupled layer responses, respectively.

For the homogeneous pairwise--pairwise duplex of
configuration~1, the uncoupled layers differ only moderately. As shown
in Fig.~\ref{fig:duplex_homogeneous}, very weak interlayer coupling
(\(\kappa_{12}=0.001\)) leaves both responses essentially unchanged for
all displayed delays. At weak-to-intermediate coupling and short delay,
the weaker layer is enhanced and the two SR curves become nearly
coincident, with only a modest reduction of the initially stronger
layer. Increasing the coupling further, especially in combination with a large delay, shifts the resonance peaks toward larger noise amplitudes and reduces their heights.

A stronger redistribution is observed in the heterogeneous
pairwise--triadic duplex of configuration~2; see
Fig.~\ref{fig:duplex_heterogeneous}. At very weak coupling, the
pairwise layer has a substantially larger resonance capacity than the
triadic layer. For weak-to-intermediate coupling and short delay, the
weaker triadic layer is markedly enhanced, while the stronger pairwise
layer is reduced, so that the two layers approach a common response
level below the uncoupled maximum of layer~1 but well above that of
layer~2.

Further increasing \(\kappa_{\mathrm{12}}\) does not continue to
improve the weaker layer. Instead, the resonance peaks shift toward
larger noise amplitudes and decrease in height, an effect amplified by
increasing \(\tau_{\mathrm{12}}\). Thus, sufficiently strong coupling combined with a large delay suppresses both layers, whereas weak-to-intermediate interlayer coupling acts primarily as a response-equalization mechanism, enhancing the initially weaker layer while generally reducing the response of the stronger one.

The remaining duplex configurations exhibit the same qualitative
pattern. Interlayer coupling tends to reduce the response mismatch
between the layers, with the extent of the improvement controlled by
the uncoupled resonance gap, whereas large interlayer delays are generally detrimental once the layers are appreciably coupled. 
A systematic comparison of all six configurations is given in the next subsection.

\subsubsection{Influence of interlayer coupling, delay, and the uncoupled resonance gap}
\label{subsubsec:results_duplex_parameters}

We compare the dependence on interlayer coupling strength and delay
across all six duplex configurations. To parameterize the initial
response mismatch, we define the uncoupled resonance-capacity gap as
\begin{equation}
\label{eq:uncoupled_resonance_gap}
\Delta Q_{\max}^{(0)}
=
Q_{\max}^{(1)}(0)
-
Q_{\max}^{(2)}(0)
>0,
\end{equation}
where \(Q_{\max}^{(\ell)}(0)\) is the capacity of layer~\(\ell\) at
\(\kappa_{\mathrm{12}}=0\). For \(\kappa_{\mathrm{12}}>0\), we denote
the coupled capacity by \(Q_{\max}^{(\ell)}(\kappa_{\mathrm{12}})\).

\begin{figure}[tbp]
\centering
\includegraphics[]{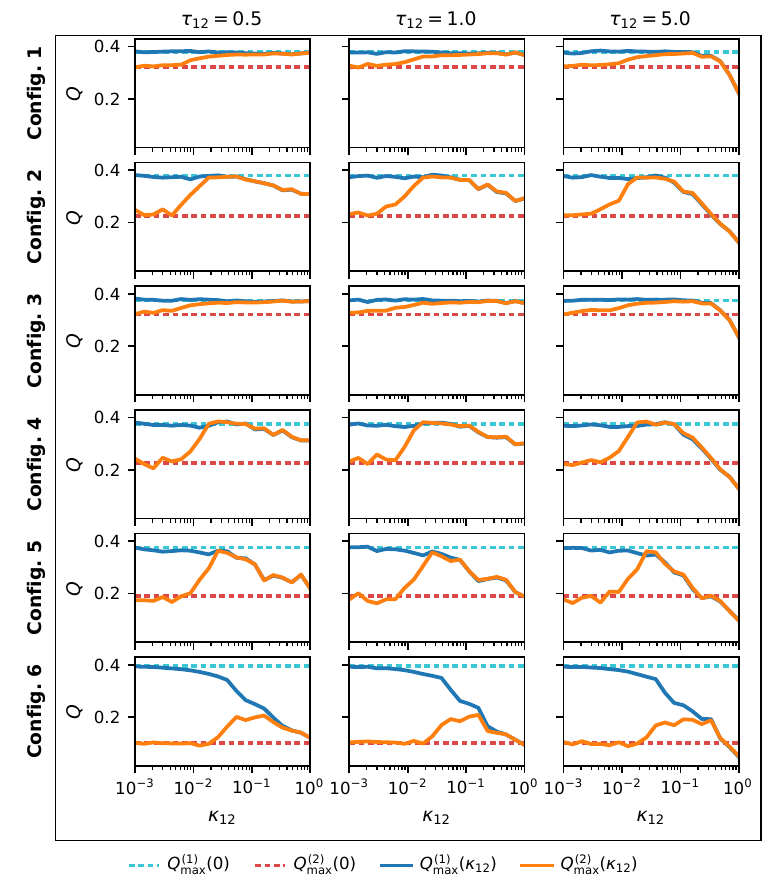}
\caption{Resonance capacities \(Q_{\max}^{(1)}\) and
\(Q_{\max}^{(2)}\) as functions of the interlayer coupling strength
\(\kappa_{\mathrm{12}}\) for \(\tau_{\mathrm{12}}=0.5\), \(1\), and
\(5\) across the six duplex configurations in
Table~\ref{tab:duplex_configurations}. Dashed horizontal lines denote
the uncoupled capacities \(Q_{\max}^{(1)}(0)\) and
\(Q_{\max}^{(2)}(0)\), whereas solid curves show
\(Q_{\max}^{(1)}(\kappa_{\mathrm{12}})\) and
\(Q_{\max}^{(2)}(\kappa_{\mathrm{12}})\).}
\label{fig:duplex_coupling_summary}
\end{figure}

Figure~\ref{fig:duplex_coupling_summary} summarizes both layer
capacities as functions of \(\kappa_{\mathrm{12}}\) for three delays.
For configurations~1 and~3, whose uncoupled gaps are comparatively
small, weak coupling rapidly equalizes the two capacities. This
near-equalized response persists over a broad weak-to-intermediate
coupling interval, particularly for short delays.

Configurations~2, 4, 5, and~6, which have larger
\(\Delta Q_{\max}^{(0)}\), exhibit a distinct coupling dependence.
Their initially weaker layer responds nonmonotonically to
\(\kappa_{\mathrm{12}}\), increasing to a maximum at
weak-to-intermediate coupling before decreasing. Hence, enhancement
of the weaker layer is optimized at a finite interlayer coupling.
Simultaneously, the stronger layer loses resonance capacity, while
coupling beyond the optimal range eventually suppresses both
responses.

Without autaptic feedback, weak-to-intermediate coupling can
substantially enhance layer~2 and, in some configurations, raise it
close to the uncoupled capacity \(Q_{\max}^{(1)}(0)\) of layer~1.
In configuration~6, which has the largest uncoupled gap and includes
autaptic feedback, layer~2 is also enhanced, but equalization occurs
at a substantially lower capacity.
Three properties of the coupling contributions account for this pattern. First, the equalizing contribution to the initially weaker layer is proportional to the mismatch it acts upon, so that at a given coupling strength it is larger when the uncoupled gap is larger. Second, this contribution is self-limiting: as the two responses approach each other, the residual mismatch, and with it the equalizing drive, decreases. Third, the instantaneous term $-\kappa_{12}v_i^{(\ell)}$ restrains both layers irrespective of the mismatch, so that it continues to grow with $\kappa_{12}$ once the equalizing contribution has saturated. For small gaps, near-equalization is therefore reached while the restraining contribution is still minor, giving a broad interval of nearly equalized responses; for larger gaps, equalization requires couplings at which the restraining contribution is already appreciable, producing an intermediate optimum of the weaker layer. The same reasoning accounts for configuration 6, in which the largest gap is associated with equalization at a substantially lower common capacity.

The equalization tendency can be understood from the
interlayer coupling acting on membrane-potential components
at the forcing frequency. At a fixed noise amplitude and parameter configuration, write these harmonic components as
\(v_{i,\Omega}^{(\ell)}(t)
=\operatorname{Re}[z_i^{(\ell)}e^{\mathrm{i}\Omega t}]\),
where \(z_i^{(\ell)}\) denotes their complex amplitude.This representation concerns only the component at
frequency \(\Omega\), not the full stochastic trajectory.
The corresponding complex amplitude
\(\widehat I_i^{(\ell)}\) of the interlayer coupling
contribution is
\begin{equation}
\widehat I_i^{(\ell)}
=
\kappa_{12}
\left[e^{-\mathrm{i}\Omega\tau_{12}}
z_i^{(3-\ell)}-z_i^{(\ell)}\right].
\end{equation}
For the difference and common amplitudes
\(z_{i,-}=z_i^{(1)}-z_i^{(2)}\) and
\(z_{i,+}=z_i^{(1)}+z_i^{(2)}\), the corresponding
coupling contributions
\(\widehat I_{i,-}=\widehat I_i^{(1)}-\widehat I_i^{(2)}\)
and
\(\widehat I_{i,+}=\widehat I_i^{(1)}+\widehat I_i^{(2)}\)
are
\begin{equation}
\widehat I_{i,-}
=-\kappa_{12}
\left(1+e^{-\mathrm{i}\Omega\tau_{12}}\right)z_{i,-},
\quad
\widehat I_{i,+}
=\kappa_{12}
\left(e^{-\mathrm{i}\Omega\tau_{12}}-1\right)z_{i,+}.
\end{equation}
At zero delay, coupling directly opposes the difference mode and has no direct contribution to the common mode. For nonzero delay, the coupling coefficients depend on the phase shift \(\Omega\tau_{12}\), and the common-mode contribution is generally nonzero. Within the sampled range this phase contribution is small, since $\Omega\tau_{12} \leq 0.1$~rad.
These relations describe the action of the coupling operator on a harmonic component; they do not constitute a closed linear-response theory for the nonlinear stochastic dynamics.

The uncoupled resonance-capacity gap alone cannot predict the location or width of the optimum: the two capacities may occur at different noise amplitudes, and their difference contains no phase or susceptibility information. Moreover, the configurations differ in their intralayer parameters as well as their gaps. The observed association therefore does not establish a universal gap-dependent scaling law. This interpretation nevertheless makes a testable prediction: at fixed interlayer delay and small $\kappa_{12}$, the normalized enhancement $[Q_{\max}^{(2)}(\kappa_{12})- Q_{\max}^{(2)}(0)]/\Delta Q_{\max}^{(0)}$ should depend only weakly on the intralayer architecture. A quantitative test of this collapse is left to future work.

\begin{figure}[p]
\centering
\includegraphics[]{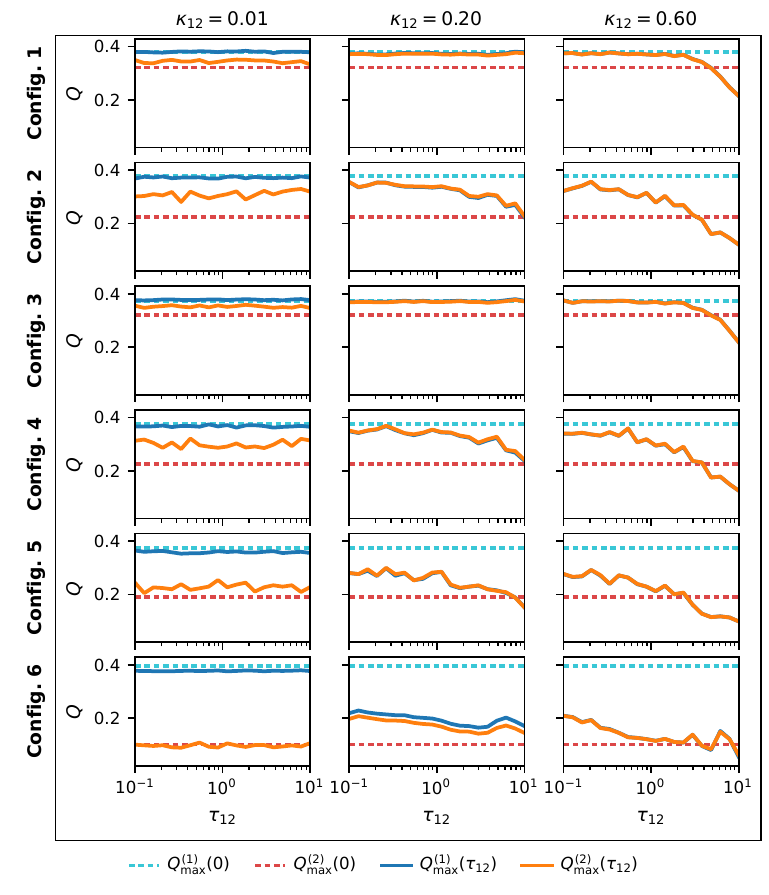}
\caption{Resonance capacities \(Q_{\max}^{(1)}\) and
\(Q_{\max}^{(2)}\) as functions of the interlayer delay
\(\tau_{\mathrm{12}}\) for \(\kappa_{\mathrm{12}}=0.01\), \(0.20\),
and \(0.60\) across the six duplex configurations in
Table~\ref{tab:duplex_configurations}. Dashed horizontal lines denote
the uncoupled capacities \(Q_{\max}^{(1)}(0)\) and
\(Q_{\max}^{(2)}(0)\), whereas solid curves show the corresponding
coupled capacities.}
\label{fig:duplex_delay_summary}
\end{figure}

The delay dependence is summarized in
Fig.~\ref{fig:duplex_delay_summary}. At very weak interlayer coupling,
varying \(\tau_{\mathrm{12}}\) has little effect because the layers
remain nearly decoupled. At intermediate and strong coupling,
increasing \(\tau_{\mathrm{12}}\) generally lowers the coupled
response, most clearly in the weaker layer and, depending on the
configuration, also in the stronger layer. The suppression is
strongest when large delays are combined with strong coupling.
Although delay can reorganize collective states in multiplex networks
\cite{triplex-chimera}, within the sampled configurations and delay range,
interlayer delay acts predominantly as a suppressive control parameter once the layers are appreciably coupled. This is an overall numerical trend, rather than strict
monotonicity at every sampled point. The investigated delays satisfy $\Omega\tau_{12}\leq 0.1$~rad, so that the relayed interlayer signal remains essentially in phase with the periodic forcing; the observed suppression therefore cannot be attributed to a delay-induced phase lag relative to the drive, and is instead consistent with a loss of coincidence between the arrival of the relayed input and the susceptible phase of the receiving neuron. Behavior at delays comparable to the forcing period $T_f$ is not addressed here.
No robust delay-induced enhancement is resolved in these scans; enhancement or other nonmonotonic behavior at different
intralayer parameters or outside the investigated delay
range is not excluded.

Although the uncoupled resonance gap \(\Delta Q_{\max}^{(0)}\) provides a useful descriptor of the observed equalization pattern, the intralayer
architecture also influences the absolute response scale.
In particular, configurations 1 and 3, whose weaker layers are weakly coupled pairwise networks and therefore have uncoupled capacities close to the isolated-neuron reference, retain the largest weaker-layer responses at long interlayer delays.

Overall, weak-to-intermediate coupling with a short interlayer delay
provides the most favorable regime for enhancing a weakly resonating layer. Strong coupling combined with long delay is generally detrimental to the SR response of both layers.

\newpage
\section{Summary and concluding remarks}
\label{sec:summary_conclusions}

We have investigated stochastic resonance in a hierarchy of delayed
FitzHugh--Nagumo systems, ranging from an isolated excitable neuron to
single-layer networks with pairwise and triadic interactions and,
finally, to duplex networks. The response was quantified by the
spectral amplitude at the forcing frequency, its maximum $Q_{\max}$
over the sampled noise amplitudes, and the corresponding optimal noise
amplitude $\sigma_{\mathrm{opt}}$. Parameter combinations producing
repetitive deterministic spiking at $\sigma=0$ were excluded from the
analysis. We have three main findings.

First, among the autapse-free single-layer architectures, pairwise and triadic interactions attain comparable maximal responses, $Q_{\max}\approx0.38$, but over different ranges of nominal coupling strength: the pairwise enhancement persists across the investigated range of $\kappa_1$, whereas the triadic enhancement is confined to weak coupling and is suppressed at larger $\kappa_2$. Under the matched participation degrees used here, the expected effective triadic coupling is twice its pairwise counterpart at equal nominal strength, so this difference is to a large extent attributable to the adopted normalization rather than to interaction order as such.

Second, delayed autaptic feedback does not appreciably increase
$Q_{\max}$ in networks containing exclusively pairwise or exclusively
triadic interactions, where it mainly shifts $\sigma_{\mathrm{opt}}$,
but, within the configurations investigated, produces a clear
enhancement of $\Delta Q_{\max}\approx0.048$, about $15\%$, when both interaction orders coexist. This architecture specificity indicates
a non-additive interplay among autaptic, pairwise, and higher-order
coupling.

Third, interlayer coupling primarily redistributes and equalizes the response between the layers: the initially weaker layer is enhanced, generally at the expense of the initially stronger one.
Within the sampled configurations, the uncoupled resonance-capacity gap \(\Delta Q_{\max}^{(0)}\) provides a useful descriptor of the observed redistribution: small gaps are associated with broad near-equalization intervals, whereas large gaps accompany more pronounced intermediate-coupling optima of the weaker layer. This association does not establish a gap-dependent scaling law, since the configurations differ in their intralayer parameters as well as their gaps. Increasing the interlayer delay predominantly suppresses the response within the sampled configurations and delay range, particularly at intermediate and strong coupling.

Taken together, the results indicate that no single coupling
mechanism is uniformly optimal. Within the investigated
configurations, interlayer coupling should be tuned according
to the uncoupled resonance gap and transmission delay.
These findings suggest potential design guidelines for
noise-assisted signal processing in delayed multilayer and
neuromorphic systems \cite{roldan2024,maldonado2023}.

The parameter maps support the reported trends over the displayed
coupling-strength and delay ranges at the specified intrinsic
and forcing parameters. Although the isolated-neuron scans
characterize the dependence on \(a\) and \(\varepsilon\),
robustness of the network-level comparisons to changes in these
parameters, the forcing parameters, or simultaneous variation
of fixed parameters remains untested. The observed suppression
with interlayer delay applies to the investigated delays,
which are short relative to the forcing period.
Delay-induced enhancement or nonmonotonic behavior at delays comparable to $T_f$ is neither observed nor excluded here.

The conclusions are further restricted to neurons with identical
intrinsic parameters, fixed network realizations, the network
size \(N=15\), and the specified linear diffusive coupling laws.
The pairwise and triadic architectures were compared under
matched expected node-participation degrees, which do not imply
equal effective coupling strengths.
Finite-size effects cannot be excluded. With fixed expected
participation and nominal coupling strengths, the \(1/N\)
normalization makes the expected intralayer coupling coefficients
decrease with network size. Assessing persistence in larger
networks therefore requires an explicit connectivity and
coupling-scaling convention.

Future work could examine larger and heterogeneous networks,
different network and hypergraph realizations, alternative
coupling normalizations, broader parameter ranges, chemical
autapses, and more general interlayer coupling schemes.
Three directions follow directly from the present analysis. A systematic comparison of the pairwise and triadic architectures at matched expected effective coupling, rather than at equal nominal strength, would isolate the effect of hyperedge-based connectivity beyond the normalization factor identified in Section~\ref{subsec:density_matching}; comparing the distribution of coupling inputs sampled at threshold crossings would further characterize the role of coincident excursions across hyperedges. A quantitative test of the normalized-enhancement collapse proposed in Section~\ref{subsubsec:results_duplex_parameters} would establish whether the uncoupled gap organizes the coupling dependence beyond the qualitative association reported here. Finally, a network-size study, with an explicit connectivity and coupling-scaling convention, would determine the persistence of the reported comparisons in larger networks.

\section*{Acknowledgements} This work was funded by the Department of Data Science (DDS), Friedrich-Alexander-Universit\"at Erlangen-Nürnberg, Germany, and the Deutsche Forschungsgemeinschaft (DFG, German Research Foundation) via the grant YA 764/1-1 to M.E.Y—Project No. 456989199.


\section*{Data availability}
The code and data supporting the findings of this study are available from the authors upon reasonable request.


\end{document}